\documentclass[12pt]{article}

\usepackage{amsmath,amssymb}
\usepackage[dvips]{graphicx}

\title{Nova Geminorum 1912 and the Origin of the Idea of 
	Gravitational Lensing}

\author{Tilman Sauer\\
	{\small Einstein Papers Project}\\
	{\small California Institute of Technology 20-7}\\
	{\small Pasadena, CA 91125, USA}\\
	{\small tilman@einstein.caltech.edu}}

\date{}

\begin{document}

\maketitle

\begin{abstract}
Einstein's early calculations of gravitational lensing, contained
in a scratch notebook and dated to the spring of 1912, are reexamined. A hitherto unknown 
letter by Einstein suggests that he entertained the idea of explaining the phenomenon
of new stars by gravitational lensing in the fall of 1915 much more seriously than
was previously assumed. A reexamination of the relevant calculations by Einstein shows
that, indeed, at least some of them most likely date from early October 1915. 
But in support of earlier historical interpretation of Einstein's notes, it is argued that
the appearance of Nova Geminorum 1912 (DN Gem) in March 1912 may, in fact, provide 
a relevant context and motivation for Einstein's lensing calculations on the occasion
of his first meeting with Erwin Freundlich during a visit in Berlin in April 1912.
We also comment on the significance of Einstein's consideration of gravitational lensing
in the fall of 1915 for the reconstruction of Einstein's final steps in his
path towards general relativity.
\end{abstract}

\section*{Introduction}

Several years ago, it was discovered that Einstein had investigated the 
idea of geometric stellar lensing more than twenty years before
the publication of his seminal note on the subject.%
\footnote{\cite{RSS97} and \cite{RS03}.} The analysis of a
scratch notebook%
\footnote{Albert Einstein Archives (AEA), call number 3-013, published
as \cite[Appendix A]{CPAE3}. A facsimile is available on
Einstein Archives Online at http://www.alberteinstein.info.}
showed that he had derived equations in notes dated 
to the year 1912 that are equivalent to those
that he would only publish in 1936.%
\footnote{\cite{Einstein1936}.} In the notes and in the paper, Einstein derived the basic lensing
equation for a point-like light source and a point-like gravitating mass.
From the lensing equation it follows readily that a terrestial
observer will see a double image of a lensed star or, in the case
of perfect alignment, a so-called ``Einstein ring.''
Einstein also derived an expression for the apparent magnification of the light
source as seen by a terrestial observer.
The dating for the notes was based on other entries in the notebook.
Some of these entries are related to a visit by Einstein in Berlin 
April 15-22, 1912, and it was conjectured that the 
occasion for the lensing entries was his meeting with
the Berlin astronomer Erwin Freundlich during this week.

The lensing idea lay dormant with Einstein until in 1936 he was prodded by the amateur 
scientist Rudi W.\ Mandl into publishing his short note in {\it Science}. In the
meantime, the idea 
surfaced occasionally in publications by other authors, such as Oliver Lodge (1919), 
Arthur Eddington (1920), and Orest Chwolson (1924).%
\footnote{\cite{Lodge1919}, \cite[pp.~133--135]{Eddington1920}, \cite{Chwolson1924}.}
We only have one other piece of evidence that
Einstein thought about the problem between 1912 and 1936. In a letter to his
friend Heinrich Zangger, dated 8 or 15 October 1915, Einstein 
remarked that he has now convinced himself that the ``new stars'' have nothing
to do with the lensing effect, and that with respect to the stellar populations
in the sky the phenomenon would be far too rare to be observable.%
\footnote{Einstein to Heinrich Zangger, 8 or 15 October 1915 \cite[Doc.~130]{CPAE8}.}

The Albert Einstein Archives in Jerusalem recently acquired a hitherto unknown letter by Einstein
that both corroborates some of the historical conjectures of the early history
of the lensing idea and also adds significant new insight into the context of Einstein's
early considerations. From this letter it appears that the phenomenon of 
``new stars,'' i.e.\ the observation of this type of cataclysmic variables, 
played a much more prominent
role in the origin of the idea than was suggested by the 
side remark in Einstein's letter to Zangger. It also adds
important new information about Einstein's thinking in the crucial period
between losing faith in the precursor theory to 
the general theory of relativity entertained in the years 1913--1915,
and the breakthrough to a general
relativistic theory of gravitation in the fall of 1915.%
\footnote{For historical discussion, see \cite{Norton1984}, \cite{ZN}, and further
	references cited therein.}
In fact, the new
letter justifies a reexamination of our reconstruction of what we know about 
Einstein's intellectual preoccupations both in April 1912 and in October 1915,
and more generally about the genesis of the concept of gravitational lensing.

\section{Einstein's letter to Emil Budde}

The new letter is a response to Emil Arnold Budde (1842--1921),
dated 22 May 1916.%
\footnote{AEA 123-079. The letter will be published in the forthcoming volume
of the {\it Collected Papers of Albert Einstein}.} Budde had been director
of the Charlottenburg works of the company of Siemens \& Halske from 1893 until 1911.%
\footnote{Budde had studied catholic
theology and science, and had worked as a secondary school 
teacher and as a correspondent for the German daily {\it K\"olnische
Zeitung} in Paris, Rome, and Constantinople. In 1887, he became
a {\it Privatgelehrter} in Berlin, edited the journal
{\it Fortschritte der Physik}, and entered the company Siemens \& Halske 
as a physicist in 1892. In 1911, he retired and moved to Feldafing, near Lake Starnberg,
since he had been advised by his physicians to live at
an altitude of at least 600m \cite{Laue1921,Werner1921}.} 
He was the author of a number
of scientific publications, among them a monograph on tensors in 
three-dimensional space \cite{Budde1914a}%
\footnote{In \cite[pp.~309--310]{Norton1992} this textbook is cited as
	evidence for the argument that Grossmann's generalization of the term `tensor'
	in \cite{EinsteinGrossmann1913} was an original development.}
and of a critical comment on relativity
published in 1914 in the {\it Verhandlungen} of the German Physical
Society.%
\footnote{\cite{Budde1914b}, \cite{Budde1914c}.}

In an unknown letter to Einstein, Budde apparently had written 
about the possibility of observing what are now called
Einstein rings, i.e.\ ring shaped images of a distant star that is in perfect 
alignment with a lensing star and a terrestial observer. 
The subject matter of Budde's initial letter can be inferred
from Einstein's response in which he pointed out that one would expect
the phenomenon to be extraordinarily rare, and that it could not be 
detected on photographic plates ``as little circles'' since irradiation
would diffuse the images that would hence only appear as bright little discs,  
indistinguishable from the image of a regular star.

The interesting part of Einstein's response follows after this negative comment. 
Einstein continued to relate that he himself had put his hopes on a 
different aspect, namely that ``due to the lensing effect''
the distant star would appear with an ``immensely increased intensity,''
and that he initially had thought that this would provide an explanation
of the ``new stars.'' He went on to list three reasons why he had given
up this hope after more careful consideration. First, the temporal development
of the intensity of a nova is asymmetric. The luminosity increases much
faster than it declines again. Second, the color of the novae usually changes
towards the red and, in general, its spectral character changes in a distinct
and characteristic way. Third, the phenomenon would be very unlikely for the
same reasons that the observation of an Einstein ring would be unlikely.
	
In the beginning of his letter, Einstein pointed out that Budde's idea concerned the same 
thing that ``about half a year ago'' (``vor etwa einem halben Jahre'')  had put him into 
``joyous excitement'' (``freudige Aufregung'').
At the end of the letter, he again wrote that the joy had been ``just as short
as it had been great.'' Counting back six months from the date of Einstein's letter,
22 May 1916, takes us to the 22nd of November 1915, which is just the time of the final 
formulation of general relativity. It is also just another six weeks or so 
away from the date of his letter to Zangger of early October, in which he wrote about the
very same subject of the possible explanation of novae as a phenomenon of gravitational
lensing.

\section{The lensing calculations in the scratch notebook}

\begin{figure}
\centering
\includegraphics[scale=0.6]{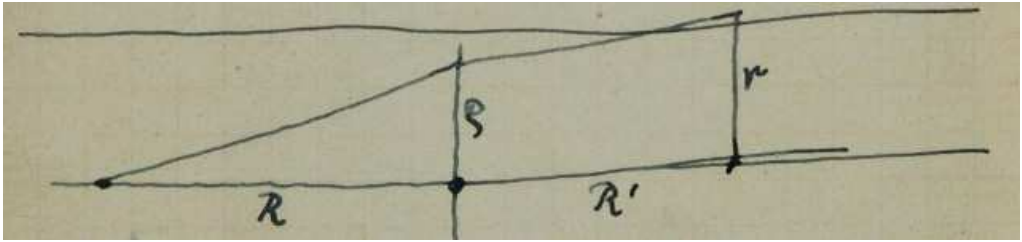}
\caption{The geometric constellation for stellar gravitational lensing as sketched in
Einstein's Scratch Notebook. From \cite[p.~585]{CPAE3}.}
\label{fig:scratch}
\end{figure}

In light of this new letter, let us briefly reexamine the calculations in the
Scratch Notebook that had been dated to April 1912.%
\footnote{The following brief recapitulation refers to \cite[585--586]{CPAE3},
or http://www.alberteinstein.info/db/ViewImage.do?DocumentID=34432\&Page=23 and $\cdots$\&Page=26. 
For a complete and detailed paraphrase of Einstein's notes, see the Appendix below.}
Stellar gravitational lensing is an implicit consequence of a law of the deflection
of light rays in a gravitational field. Such a law had been obtained by Einstein in 1911
as a direct consequence of the equivalence hypothesis.
The angle of deflection $\tilde{\alpha}$%
\footnote{I am using the notation $\tilde{\alpha}$ instead of $\alpha$ (as in \cite{Einstein1911}) 
in order to distinguish this angle
from the quantity $\alpha$ (effectively the Schwarzschild radius) in Einstein's scratch notebook.}
was found to be 
\begin{equation*}
\tilde{\alpha} = \frac{2kM}{c^2\Delta},
\end{equation*}
where $k$ is the gravitational constant, $M$ the mass of the lensing star, $c$ the
speed of light, and $\Delta$ the distance of closest approach
of the light ray measured from the center of the massive star.%
\footnote{\cite[p.~908]{Einstein1911}. Qualitatively, Einstein had already derived the
consequence of light bending in a gravitational field when he first formulated his
equivalence hypothesis \cite[p.~461]{Einstein1907}. In the final theory of general
relativity, the same relation is obtained with an additional factor of $2$, as observed
explicitly in \cite[p.~834]{Einstein1915c}. Incidentally,
the relevant formula was printed incorrectly by a factor of $2$ in (the first printing of)
Einstein's 1916 review paper of general relativity \cite[p.~822]{Einstein1916}, see 
\cite[Doc.~30, n.~36]{CPAE6} and also Einstein's
response to Carl Runge, 8 November 1920 \cite[Doc.~195]{CPAE10}.}
On [p.~43] of the Scratch Notebook we find the sketch shown in Fig.~(\ref{fig:scratch})
and underneath it the lensing equation
\begin{equation*}
r = \rho\frac{R+R'}{R} - \frac{R'\alpha}{\rho},   
\end{equation*}
where $R$ denotes the distance between the light emitting distant star and the massive
star that is acting as a lens, $R'$ the distance between the lensing star and the position of 
a terrestial observer who is located a distance $r$ away from the line connecting light source
and lensing star. $\rho$ is the distance of closest approach of a light ray emitted by the
star and seen by the observer. $\alpha=2kM/c^2$ is a typical
length (later known as the Schwarzschild radius) that depends on the mass of the light
deflecting star and that determines the angle of deflection to be $\frac{\alpha}{\rho}$.
The lensing equation can be written in dimensionless variables as
\begin{equation}
r_0 = \rho_0 - \frac{1}{\rho_0},
\label{eq:lensingdimless}
\end{equation}
after defining $r_0$ and $\rho_0$ as 
\begin{eqnarray}
r_0 &=& r\sqrt{\frac{R}{R'(R+R')\alpha}}, \notag \\
\rho_0 &=& \rho \sqrt{\frac{R+R'}{RR'\alpha}}. 
\end{eqnarray}
The fact that equation (\ref{eq:lensingdimless}) is a quadratic equation for $\rho_0$
entails that there are two solutions which correspond to two light rays that can reach 
an observer, along either side of the lensing star,%
\footnote{Since only three points are given, the problem is intrinsically a planar one,
	as long as the three points are not in perfect alignment.}
	and hence that a terrestial observer will see a double image of the
distant star. For perfect alignment, the double image will turn into a ring shaped image, 
an ``Einstein-ring'' whose diameter 
$\rho^{\rm ring}_0=\rho^{\rm ring}\sqrt{\frac{R+R'}{RR'\alpha}}=1$ also follows immediately
from the lensing equation.

In light of Einstein's letters to Zangger and Budde, it is interesting that  
Einstein went on to compute also the apparent magnification, obtaining the following 
expression:
\begin{equation}
\mathcal{H}_{\rm tot} = H\left\{ \frac{1}{1-\frac{1}{\rho_1^4}} +
\frac{1}{\frac{1}{\rho_2^4}-1}\right\}.
\end{equation}
Here $\mathcal{H}_{\rm tot}$ is the total intensity received by the observer, and
$H$ the intensity of the star light at distance $R$. $\rho_{1,2}$ denote the two 
roots of the quadratic equation (\ref{eq:lensingdimless}).
The term in brackets gives the relative brightness, reducing to $1$ if no lensing
takes place.
Finally, some order of magnitude calculations on these pages showed that the probability of
observing this effect would be given by the probability of having two stars within a solid angle
that would cover $10^{-15}$ of the sky, which is highly improbable given that the number of known
stars at the time was of the order of $10^6$.%
\footnote{See the discussion in the appendix.}

Equations that are
entirely equivalent to these were published much later, in 1936, in Einstein's
note to {\it Science}.%
\footnote{\cite{RSS97}.}

The dating of the lensing notes in the scratch notebook to Einstein's visit
in Berlin in April 1912 was based on other evidence in the notebook.
Most importantly, p.~[36] lists Einstein's appointments during his
Berlin visit. In addition, pp.~[38] and [39] recapitulate very specifically the 
equations of Einstein's two papers on the theory of the static gravitational
field of February and March 1912, respectively.%
\footnote{\cite{Einstein1912a,Einstein1912b}.}
The calculations that deal with the lensing problem then appear on 
pp.~[43]-[48], and on pp.~[51] and [52] of the notebook. The sheet 
containing pp.~[44] and [45] is a loose sheet inserted between 
p.~[43] and p.~[45]. After p.~[53], three pages have been torn out, and then 
follow 37 blank pages, with some pages torn out in between. The remainder
of the notebook contains entries that begin at the other end of the notebook
which was turned upside down.
Except for some apparently unrelated and undated entires on pp.~[49], [50],%
\footnote{On the bottom half of p.~[49] there is a sketch of Pascal's and
	Brianchon's Theorems, which deal with hexagons
	inscribed in or circumscribed on a conical section. 
	I wish to thank Jesper L\"utzen for this identification. Other entries
	on pp.~[49] and [50] also appear to deal with problems from projective
	geometry. There is also a sketch of a vessel filled with a liquid and the
	words ``eau glycerin\'e'' and what appears to be sketch of a magnetic
	moment in a sinusoidal magnetic field.}
and [54], the lensing calculations hence are at the end of a
more or less continuous flow of entries. These physical characteristics
of the notebook lead to an important consequence. All information that was pointing to 
a date of the lensing calculations in the year 1912 preceded the actual
lensing calculations. Reexaming pp.~[51] and [52] of the notebook
in light of the letters to Zangger and to Budde in fact reveals that
at least these entries were not written in 1912, but rather
most likely at the time of the letter to Zangger, in early 
October 1915. There are two reasons for this. First, at the top of
p.~[51], Einstein wrote down the title of a book published only in 1914.%
\footnote{\cite{Fernau1914}. Could it be that the book was mentioned to Einstein when
he met with Romain Rolland in Geneva in September 1915, see \cite[Doc.~118]{CPAE8}?}
Therefore, the following calculations are almost certainly to be dated later 
than the publication
of this book. Second, at the bottom of p.~[52], Einstein explicitly
refers to the ``apparent diameter of a Nova st[ar].'' The calculations
on pp.~[51] and [52] in fact are a calculation of the apparent brightness
and diameter of a star. We conclude that, in all probability,
the calculations on pp.~[51] and [52] were written at the time of Einstein's
letter to Zangger, early October 1915.

Does the dating of pp.~[51] and [52] to October 1915 also compel us 
to revise our dating of the other lensing calculations in the notebook? 
To answer this question, we need to consider the broader historical context
of the notes. But before doing so,
we first observe that pp.~[49] and [50] contain entries that appear
unrelated to the lensing problem. As shown by the detailed paraphrase given 
in the appendix, the calculations on pp.~[43] to
[48] on the other hand represent a coherent train of thought, as do the
calculations of pp.~[51] and [52]. We also note that Einstein
used a slightly different notation on pp.~[43]ff.\ and on pp.~[51]-[52].
In the first set, he denoted the distances between light source
and lens and between lens and observer as $R$ and $R'$, respectively. On 
pp.~[51]-[52] he used the notation $R_1$ and $R_2$, respectively. He also
reversed the roles of $r$ and $\rho$.
We conclude that there is a discontinuity between the first
set of lensing calculations on pp.~[43] to [48] and the second set
on pp.~[51] and p.~[52].

\section{The context of Einstein's early lensing calculations}

From Einstein's letter to Budde we learn that he had investigated the idea
that stellar lensing might explain the phenomenon of the ``new stars,''
and that he had given up this idea after looking more closely into the
characteristic features of novae, especially their light curves and the 
changes in their spectral characteristics. Let us therefore
briefly look into the astronomical knowledge about novae at the time.

The observation of a new star is an event that, in the early twentieth century,
occurred only every few years. Between 1900 and 1915, eight novae were observed:%
\footnote{For the following, see \cite{Duerbeck1987}.}
Nova Persei 1901 (GK Per), Nova Geminorum (1) 1903 (DM Gem), Nova Aquilae
1905 (V604 Aql), Nova Vela 1905 (CN Vel), Nova Arae 1910 (OY Ara), Nova Lacertae 1910 (DI Lac), 
Nova Sagittarii 1910 (V999 Sgr), and Nova Geminorum (2) 1912
(DN Gem) with maximum brightness of 0.2, 4.8, 8.2, 10.2, 6.0, 4.6, 8.0, 
3.5 magnitudes, respectively.
At the time, ``the two most interesting Novae of the present century,''
\cite[p.~493]{Campbell1914}, were Nova Persei of 1901 and Nova
Geminorum of 1912. The next spectacular nova to occur was the very bright Nova 
Aquilae 1918 (V603 Aql) with a maximum brightness of $-1.1$~mag. 

\begin{figure}
\includegraphics[scale=0.75]{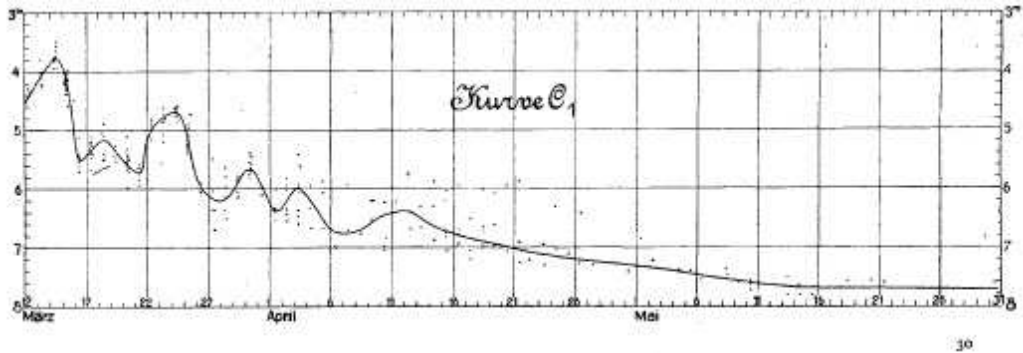}
\caption{The light curve of Nova Geminorum 1912 for the first three months
	after its appearance, as put together by Fischer-Petersen on 
the basis of 253 individual observations. The points are the magnitudes reported by
the individual observers, the solid line is to guide the eye.
From \cite[p.429]{FischerPetersen1912}.}
\label{fig:lightcurve}
\end{figure}

Nova Geminorum (2) was discovered on March 12, 1912, by the astronomer
Sigurd Enebo at Dombaas, Norway \cite{Pickering1912a}.
On a photographic plate taken at
Harvard College Observatory on March 10, showing stars of magnitude 10.5,
it was not visible, but it was visible as a magnitude 5 star in the constellation
Gemini on a Harvard plate
of March 11. On March 13, a cablegram was received at Harvard and distributed 
throughout the United States. In the following days all major observatories
as well as many amateur astronomers
pointed their instruments towards the new star. The maximum brightness of mag 3.5 was
reached on March 14 (Einstein's 33rd birthday!) \cite{FischerPetersen1912}.
By March 16, the brightness was down to a magnitude of 5.5 and in the following
weeks it decreased further, with distinct oscillations. By mid-April 1912,
most observers registered a brightness of mag $6\approx 7$,
see Fig.~(\ref{fig:lightcurve}).
We now know that the DN Gem is a fast nova with a $t_3$-time of 37d. Its light curve is type 
Bb in the classification of \cite{Duerbeck1987}, i.e.\ it declines with major
fluctuations. 

Like all classical novae, Nova Geminorum is, in fact, a binary system of a white dwarf and 
main sequence star, where hydrogen-rich matter is being accreted onto the white dwarf.
Recent observations have even determined the binary period \cite{RetterEtAl1999}.
The eruption of a classical nova occurs when a hydrogen-rich envelope of the white dwarf
suffers a thermonuclear runaway.%
\footnote{For a review, see \cite{Shara1989}.} 
This explanation of classical novae also entails that they display the same sequence of spectral
behaviour as the luminosity decreases, see also Fig.~(\ref{fig:spectra}) below.
\begin{figure}
\centering
\includegraphics[scale=0.75]{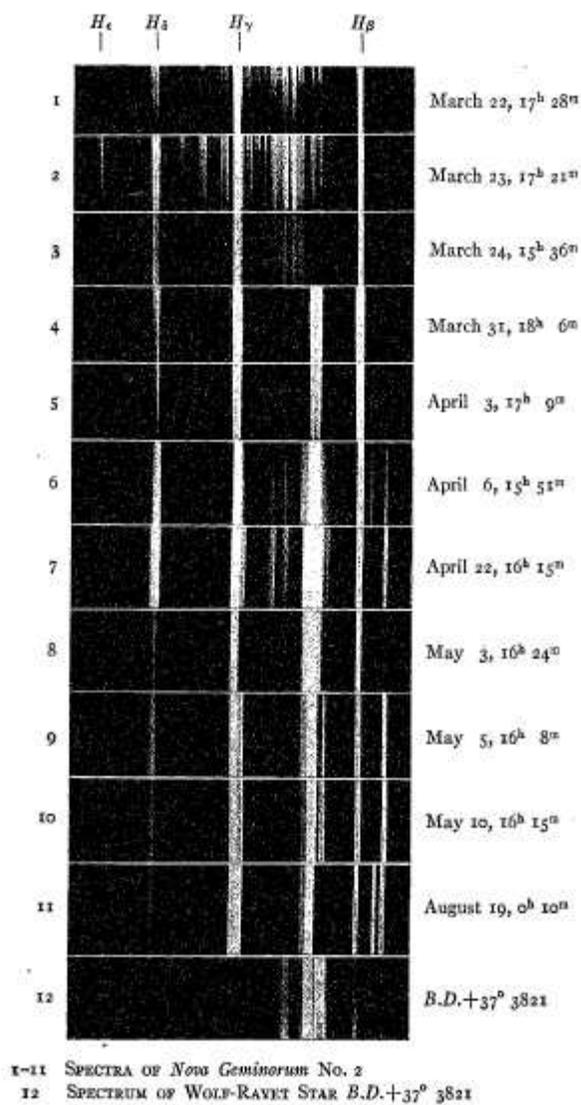}
\caption{Changes in the spectrum of Nova Geminorum 1912, March 22 to August 19, 1912. 
From \cite{AdamsKohlschutter1912}.}.
\label{fig:spectra}
\end{figure}
However, our current understanding of classical novae was suggested only in the fifties.%
\footnote{For a historical overview of previous theories, see \cite{Duerbeck2007}.}

The temporal proximity of the appearance of Nova Geminorum 1912 with Einstein's
Berlin visit during the week of April 15--22, suggests that this astronomical
event was discussed also when Einstein met with Freundlich for the first time.%
\footnote{For evidence that Einstein met with Freundlich, see his letter
	to Michele Besso, 26 March 1912, in which he mentions planned discussions 
	(``Besprechungen'') with Nernst, Planck, Rubens, Warburg, Haber,
	and ``an astronomer''---presumably Freundlich \cite[Doc. 377]{CPAE5}.}
We know that the observatory in Potsdam took a number of photographs
of the new star between March 15 and April 12 \cite{Furuhjelm1912,Ludendorff1912},
and that Freundlich, among others, was charged with photometric observations
of the nova \cite[p.~429]{FischerPetersen1912}. Einstein and 
Freundlich had earlier corresponded
about the possiblity of observing gravitational light deflection through the
gravitational field of the sun.%
\footnote{Einstein to Freundlich, 1 September 1911, 21 September 1911, and 8 January 1912 
\cite[Docs.~281, 287, 336]{CPAE5}.}
The purpose of their meeting was to
discuss possible astronomical tests of Einstein's emerging relativistic theory
of gravitation. The recent observation of the brightest
nova since 1901 must have been on Freundlich's mind, and it seems
more than likely that the idea of explaining the phenomenon in terms of
gravitational lensing therefore came up in the course of their conversation.
We conclude that our earlier dating of the first set of calculations of the lensing
problem in the Scratch Notebook to the time of Einstein's encounter with 
Freundlich in April 1912 is the most likely possibility. 

In fact, the context of the observation of Nova Geminorum 1912 provides an answer to the
question as to why Einstein would have done the calculations at all and, in particular, why
he would not have been content at the time with a
calculation of the lensing equation, the separation of the double star image and, perhaps, 
the radius of the Einstein ring.
Without this context it might seem a rather ingenious move on Einstein's part
to go ahead and immediately compute the apparent magnification of the lensed star as well.
But this answer to the question of motivation for the specific details of
the calculations in the Scratch Notebook, immediately raises another question.

Assuming that the first set of lensing calculations were done in spring 1912, why do
we have no evidence that this idea was followed up by either Einstein or by Freundlich until
the fall of 1915? To answer this question, it should first be observed that
no summarizing results and analyses of the observations of Nova Geminorum 1912 were published
before the end of the summer.

Let us briefly recall Einstein's intellectual preoccupations after his visit to Berlin
in April 1912.%
\footnote{We will focus here on his work of gravitation yet for the sake of completeness
	it should be noted that Einstein at the same time was also thinking about
	quantum theory, most notably about the law of photochemical
	equivalence and about the problem of zero point energy, see \cite[Docs.~5, 6, 11, 12]{CPAE4}.}
Shortly before his trip to Berlin he had submitted his two papers on a theory of the
static gravitational field.%
\footnote{\cite{Einstein1912a}, \cite{Einstein1912b}, were received by the {\it Annalen der
	Physik} on 26 February and 23 March, respectively.}
After his return to Prague in April 1912, Einstein was preparing for his move to
Zurich. The two papers were published in the 23 May issue of the {\it Annalen der
	Physik}. Einstein wrote an addendum at proof stage to the second one, in which he
showed that the equations of motion could be written in a variational form, adding that this
would give us ``an idea about how the equations of motion of the material point in a dynamic
gravitational field 
are constructed'' \cite[p.~458]{Einstein1912b}. He also entered into a published dispute
with Max Abraham on their respective theories of gravitation.%
\footnote{\cite{Einstein1912c} which was received by the {\it Annalen} on 4 July 1912 is a response
	to a critique by Abraham.}
At the end of July, he departed Prague for Zurich.
The next thing we know about his work on gravitation comes from a letter to Ludwig Hopf, 
dated 16 August 1912, in which he wrote:
\begin{quote}
The work on gravitation is going splendidly. Unless I am completely wrong, I have now found the
most general equations.%
\footnote{Einstein to Hopf, 16 August 1912 \cite[Doc.~416]{CPAE5}.}
\end{quote}
These most general equations are, in all probability, equations of motion in a gravitational field,
represented by a metric tensor. After his arrival in Zurich,
Einstein began a collaboration with his former classmate Marcel Grossmann, now his
colleague at the ETH. Their research on a generalized theory of relativity is documented
in Einstein's so-called ``Zurich Notebook''%
\footnote{AEA~3-006, see \cite[Doc.~10]{CPAE4}. For a comprehensive discussion of this document,
including a facsimile, transcription, and detailed paraphrase, see \cite{ZN}.}
and culminates in the publication of the ``Outline [{\it Entwurf}] of a generalized 
theory of relativity and a theory of gravitation,'' in early summer of 1913 
co-authored with Marcel Grossmann.%
\footnote{\cite{EinsteinGrossmann1913}.} 
This so-called {\it Entwurf}-theory contains all the elements of the final
theory of general relativity, except for generally relativistic field equations. Einstein
would hold onto this theory until his final breakthrough to general relativity in the fall 
of 1915.

In conclusion, we observe that Einstein's path toward the general theory of relativity in 1912
took him deep into the unknown land of the mathematics associated with the metric tensor,
before there was a chance to reconsider the lensing idea in light of the data for 
Nova Geminorum 1912. In any case, he would have to rely on Freundlich or other professional
astronomers for a secure assessment of the possibilities of an observation of the lensing 
effect at the time.

Freundlich, on the other hand, continued to think about ways to test Einstein's new theory of 
gravitation.%
\footnote{See \cite{Hentschel1994} and \cite{Hentschel1997}.}
But his focus was on observations of light deflection during a solar eclipse.%
\footnote{See his correspondence with Einstein in \cite{CPAE5}.} In August 1914, he led a
first (unsuccessful) expedition to the Crimea to observe the eclipse of 21 August 1914. 
Even these efforts were hampered by the lack of funding and, more generally, by the 
difficulties of securing
increased research time that would have allowed Freundlich to freely pursue
his collaboration with Einstein.

Given these circumstances, and the fact that order-of-magnitude calculations may have convinced
Einstein already in 1912 that the phenomenon would be rare, it seems plausible that 
the lensing idea was not pursued further for some time after Einstein's visit in Berlin in
April 1912.

Let us finally reexamine the events of fall 1915. Einstein, in the meantime had left Zurich
in the spring of 1914, accepting an appointment as member of the Prussian Academy in Berlin. 
In September 1915, Einstein spent
a few weeks in Switzerland where he met, among others, with Heinrich Zangger, Michele Besso,
and Romain Rolland. On 22 September 1915, he left Zurich%
\footnote{\cite[p.~998]{CPAE8}.}
but travelled via Eisenach where he was on the 24th of September.%
\footnote{\cite[Doc.~Vol.~8, 122a]{CPAE10}.}
By the 30th of September, at the latest, he was back in Berlin, and wrote a letter to Freundlich:
\begin{quote}
I am writing you now about a scientific matter that electrifies me enormously.%
\footnote{Einstein to Freundlich, 30 September 1915 \cite[Doc.~123]{CPAE8}. For a detailed
	discussion of this letter and its significance for the reconstruction of
	Einstein's final breakthrough to general relativity, see \cite{Janssen1999}.}
\end{quote}
It is clear from the letter, however, that the excitement indicated to Freundlich is {\it not}
about the idea of gravitational lensing. Rather, Einstein had found an internal contradiction
in his {\it Entwurf} theory that amounted to the realization that Minkowski space-time
in rotating Cartesian coordinates would not be a solution of the {\it Entwurf} field equations.%
\footnote{Interestingly, the Scratch Notebook contains an entry that is pertinent
	to this problem. On p.~[66], i.e.\ on the last page of the backward end of the notebook,
	Einstein considers the case of rotation in a calculation that exactly matches corresponding
	calculations dating from October 1915, see \cite{Janssen1999}. Janssen cautiously remarks
	that he believes this calculation to date from 1913 \cite[p.~139]{Janssen1999}. It seems 
	possible, however, that
	these entries as well as the immediately preceding ones on the perihelion advance
	(see note \ref{note:perihelion}) may well
	date from late 1915 as well. \label{note:rotation}}
This insight undermined his confidence in the validity of the {\it Entwurf} theory, and is later 
mentioned as one of three arguments that induced Einstein to lose faith in the {\it Entwurf}
equations.%
\footnote{See Einstein to Arnold Sommerfeld, 28 November 1915, and to Hendrik A.~Lorentz, 1 January
	1916 \cite[Docs.~153, 177]{CPAE8}.}
The first of these arguments was the fact that a calculation of the
planetary perihelion advance in the framework of the {\it Entwurf} theory did not
produce the well-known anomaly that had been established for Mercury. This problem had been 
known to Einstein for some time.%
\footnote{See \cite{EarmanJanssen1993} and \cite[pp.~344--359]{CPAE4}. The Scratch Notebook contains
some calculations related to the perihelion advance on pp.~[61--66], i.e.\
in the backward end of the notebook. On p.~[61], Einstein there explicitly noted that the advance
of Mercury's perihelion would be $17''$ which is the value that is obtained on the basis of the
{\it Entwurf}-theory. These calculations are undated, see note \ref{note:rotation}.
\label{note:perihelion}}
The third argument was realized sometime in early October, a few days after stumbling upon the problem
with rotation, and concerned the mathematical derivation of the 
{\it Entwurf} field equations in Einstein's comprehensive review of October 1914.%
\footnote{\cite{Einstein1914}.}
In any case, we know that Einstein asked Freundlich to look into the problem of the rotating 
metric, and that they met some time in early October. This follows from a letter Einstein
wrote to Otto Naumann, dated after 1 October 1915, in which Einstein asked about possibilities
to allow Freundlich more freedom to pursue independent research. In this letter, Einstein mentioned
that Freundlich had visited him ``recently.''%
\footnote{``Letzter Tage war Herr Dr.~Freundlich von der Sternwarte N bei mir.'' 
	\cite[Doc.~124]{CPAE8}.}

By 12 October, Einstein had realized the third problem with the {\it Entwurf} theory, the 
unproven uniqueness of the Lagrangian for the {\it Entwurf} field equations, as he reported in a letter
to Lorentz. In this letter, he neither mentioned the problem with the rotating metric nor the
issue of gravitational lensing.%
\footnote{In a letter to Hilbert, dated 7 November 1915, Einstein wrote that he realized the
flaw in his proof ``about four weeks ago'' \cite[Doc.~136]{CPAE8}.} 

For our reconstruction of this episode, the precise date of Einstein's letter to Zangger in
which he remarked that he had given up the hope of explaining the ``new stars'' as a lensing
phenomenon is relevant. It could have been written either on the 8th or the 15th 
of October.%
\footnote{The editors of \cite{CPAE8} dated this letter explicitly to the 15th of  
October. It seems, however, that the 8th is also a possibility. The letter was written on a
Friday between September 30, when a fire and explosion took place in the comb factory Walter
near Lake Biel took place, mentioned in the letter, and October 22 when Einstein
participated in the first Academy session after the summer break. I see no reason
why Einstein could not have heard of the accidents from Zangger before October 8.}

The letter to Zangger suggests that they had talked about the idea earlier since Einstein seems
to presuppose that Zangger knew what he was talking about and did not
explain what he meant by ``lens effect'' (``Linsenwirkung''). 
As mentioned before, Einstein had just recently met with 
Zangger, as well as with Besso before returning to Berlin. The following scenario seems therefore
plausible:

Upon returning to Berlin some time after the 24th of September 1915, Einstein realized the problem of the
rotating metric solution and wrote to Freundlich on the 30th, asking him to look into this issue.
Shortly afterwards, the two met in person. Most likely they discussed not only the
rotation problem, but also the lensing idea. Having found troubling indications of an inner inconsistency
in the very foundations of this theory, it would have been a natural move for Einstein
to go back and reconsider early arguments such as one based safely on the equivalence hypothesis.%
\footnote{It seems unlikely that Einstein at that time was already contemplating a quantitatively
	different law of light deflection. Einstein first observed in \cite[p.~834]{Einstein1915c}
	that an additional factor of $2$ would arise from the different first-order approximation for
	the metric if the Newtonian limit is derived on the basis of generally
	covariant field equations in which the Ricci tensor is directly set proportional 
	to the energy-momentum tensor. These latter equations were published in his second November
	memoir, presented on 11 November, under the assumption that the trace of the energy-momentum
	tensor vanish. In his comment on the factor of $2$, Einstein refers to this result as being in
	contrast to ``earlier calculations'' 
	where the hypothesis of vanishing energy-momentum had not yet been made.}
After this meeting, Einstein wrote to
Naumann exploring possibilities to give Freundlich more research freedom. By October 8,
Einstein had convinced himself that gravitational lensing cannot explain the ``new stars.''
On 12 October, he realized the third problem of his mathematical derivation of the 
{\it Entwurf} field equation.

According to this reconstruction of the sequence of events, it is remarkable that the ``joyous excitement''
about the lensing idea falls within days after his being ``electrified'' about the realization
of the rotation problem on 30 September, and his realization of the third problem of the 
mathematical derivation of the {\it Entwurf} equation, on or before 12 October 1915.%
\footnote{For completeness, one should point one other intellectual activity
	of Einstein's during those days. In Einstein's letter to Zangger of 8 or 15 October,
	he also mentioned that he wrote ``a supplementary paper to my last year's analysis on
	general relativity.'' The last year's analysis is, in all likelyhood \cite{Einstein1914};
	the supplementary paper is, in all likelihood, an early version of \cite{Einstein1916b},
	or, perhaps, an early version of Einstein's first November memoir \cite{Einstein1915a},
	see \cite[Doc.~130, note~5]{CPAE8} and \cite[note~51]{Janssen1999}.}

Some five weeks later, his excitement was even greater and his heart, allegedly, skipped 
a beat when he found that
he could derive the anomalous advance of Mercury's perihelion on the basis of his new
field equations. And after having submitted the last of his four November communications 
to the Prussian Academy on 25 November which presented the final gravitational field equations, 
the ``Einstein equations,'' he wrote to Sommerfeld:
\begin{quote}
You must not be cross with me that I am answering your kind and interesting letter only today.
But in the last month I had one of the most exciting, exhausting times of my life, indeed 
also one of the most successful. I could not think of writing.%
\footnote{``Sie d\"urfen mir nicht b\"ose sein, dass ich erst heute auf Ihren freundlichen 
	und interessanten Brief antworte. Aber ich hatte im letzten Monat eine der aufregendsten,
	anstrengendsten Zeiten meines Lebens, allerdings auch der erfolgreichsten.'' 
		Einstein to Sommerfeld, 28 November 1915 \cite[Doc.~153]{CPAE8}.}
\end{quote}
It is interesting to learn from Einstein's letter to Budde that in addition to the realization of
the problems with the {\it Entwurf} theory and the eventual success of 
his breakthrough to general relativity, an astronomical problem, the idea of explaining novae 
in terms of gravitational lensing added to Einstein's excitement in the midst of what must 
indeed have been the most intense period of intellectual turmoil in his life.

\section{Concluding remarks}

Einstein's recollections of his thought concerning the explanation of the ``new stars'' 
as a phenomenon of gravitational lensing in his letter to Budde add two significant insights to 
our reconstruction of the genesis
of general relativity. If our dating and context hypothesis of the lensing calculations
in the scratch notebook are correct, we learn that it was an astronomical observation
that triggered the elaboration of a significant consequence of the equivalence hypothesis
and its consequence of gravitational light deflection. It is also interesting that on his
intellectual path from the {\it Entwurf} theory to the final theory of
general relativity, Einstein also took a detour in which he explored further consequences of one of 
the solid pillars of general relativity, the equivalence hypothesis.

\section*{Appendix: Einstein's lensing calculations in the Scratch Notebook AEA 3-013}

The following is a self-contained line-by-line paraphrase of Einstein's lensing calculations in
his scratch notebook, \cite[pp.~585--589]{CPAE3}. The pagination in square
brackets refers to the sequence of pages in the notebook.

The calculations start out on p.~[43] with Fig.~(\ref{fig:scratch}) and continue on the
facing page p.~[46].
From the more explicit sketch in Fig.~(\ref{fig:geometry}), we read off the lensing equation:
\begin{figure}
\centering
\includegraphics[scale=0.5]{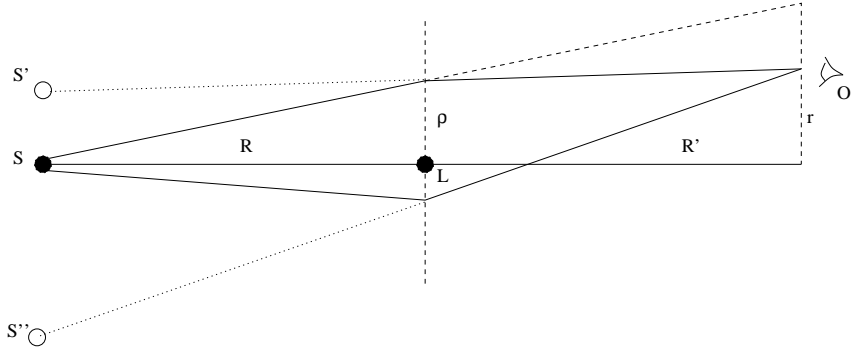}
\caption{The geometry of stellar lensing.}
\label{fig:geometry}
\end{figure}
\begin{equation}
r = \rho\frac{R+R'}{R} - \frac{R'\alpha}{\rho}. 
\label{eq:app:lensing}
\end{equation}
Here $R$ is the distance between the light emitting star $S$ and the lensing star $L$;
$R'$ the distance between the massive star $L$ and the projected position of the observer $O$
on the line connecting light source and lens; $\rho$ is the distance of closest approach of 
a light ray emitted from the distant star and seen by an observer; $r$ is the orthogonal
distance of the terrestial observer to the line connecting light source and lens. The first term 
in the lensing equation (\ref{eq:app:lensing}) is obtained from the similarity
of triangles with baseline $R$ and $R+R'$, respectively, and the second term is the angle of
deflection as given by the law of gravitational light bending, where $\alpha$ is the Schwarzschild
radius $2GM/c^2$.
If we want to write this equation in dimensionless variables,
we need to multiply it by a factor of
\begin{equation}
\sqrt{\frac{R}{R'(R+R')\alpha}}
\label{eq:app:factor}
\end{equation}
so that, when we define $r_0$ and $\rho_0$ as
\begin{eqnarray}
r_0 &=& r\sqrt{\frac{R}{R'(R+R')\alpha}} \label{eq:app:r0} \\
\rho_0 &=& \rho \sqrt{\frac{R+R'}{RR'\alpha}}  
\end{eqnarray}
the lensing equation (\ref{eq:app:lensing}) turns into
\begin{equation}
r_0 = \rho_0 - \frac{1}{\rho_0}.
\label{eq:app:lensingdimless}
\end{equation}
This is a quadratic equation for $r_0$, the two solutions of which
correspond to the two light rays passing above and below $L$. The observer $O$
therefore sees two images of $S$ at positions $S'$ and $S''$, respectively.
To read off the radius of an ``Einstein ring,'' obtained for perfect alignment of
$S$, $L$, and $O$, one only needs to set $r_0\equiv 1$.

In order to get an expression for the apparent magnification, Einstein proceeded as follows.
He first took the square of eq.~(\ref{eq:app:lensingdimless}) as
\begin{equation}
2+r^2 = \rho^2 + \frac{1}{\rho^2}.
\end{equation}
If we multiply this equation by $\pi$ and denote the areas of the circles corresponding
to the radii $r$ and $\rho$ as $f=\pi r^2$  and $\varphi=\pi \rho^2$, respectively, we
can write this equation as 
\begin{equation}
2\pi + f = \varphi + \frac{\pi^2}{\varphi}.
\end{equation}
We are not interested in the full circle corresponding to these radii but in the differential
area element associated with these radii. More precisely, we are interested in the change
of the differential area element $df$ associated with $f$ when we change the differential area
element $d\varphi$ associated with $\varphi$. Hence, Einstein wrote
\begin{equation}
df = \left(1-\frac{\pi^2}{\varphi^2}\right)d\varphi  = \left(1-\frac{1}{\rho^4}\right)d\varphi.
\label{eq:app:df}
\end{equation}
The intensity $\mathcal{H}$ of the brightness received at $r$ is related to the
intensity $H$ of the brightness at $\rho$ by
\begin{equation}
\mathcal{H}df = \pm H d\varphi,
\end{equation}
where the plus and minus signs refer to the two solutions of the quadratic equation.
Since we have from (\ref{eq:app:df}) 
\begin{equation}
\frac{df}{d\varphi} = \left(1-\frac{1}{\rho^4}\right), 
\end{equation}
we get
\begin{equation}
\mathcal{H} = \pm \frac{H}{1-\frac{1}{\rho^4}}.
\end{equation}
or, inserting the explicit solutions, we can write the total
brightness at $r$ as
\begin{equation}
\mathcal{H}_{\rm tot} = H\left\{ \frac{1}{1-\frac{1}{\rho_1^4}} +
\frac{1}{\frac{1}{\rho_2^4}-1}\right\}.
\end{equation}
As Einstein remarked, the term in brackets gives the relative brightness, if we take
the value for $r\rightarrow\infty$ to be $1$.%
\footnote{``Klammer gibt relative Helligkeit''}
This result is equation number (3) in Einstein's notes, and most of the following
material on pp.~[47] and [48], as well as on the loose sheet containing pp.~[44] and [45],
will be a discussion of this expression for the relative brightness. 

On p.~[47], Einstein first rewrote the reduced lensing equation as
\begin{equation}
r = \frac{1}{x} - x,
\label{eq:lensing3}
\end{equation}
and then the terms in brackets as
\begin{equation}
\left\{ \right\} = \frac{1}{1-x_1^4} + \frac{1}{x_2^4-1}.
\end{equation}
The next step is to bring the two terms to a common denominator%
\footnote{In the notes, Einstein refers to this step as ``Rationalisierung''.}
\begin{equation}
\mathcal{H}_r = \frac{x_1^4 - x_2^4}{(1-x_1^4)(1-x_2^4)}.
\label{eq:app:Hr}
\end{equation}
If one squares the lensing equation (\ref{eq:lensing3}) twice, one obtains
\begin{equation}
-2+(2+r^2)^2 = \frac{1}{x^4} + x^4.
\label{eq:app:quad}
\end{equation}
If we now introduce new variables $A$ and $u$ via
\begin{equation}
2A = -2 + (2+r^2)^2,
\end{equation}
or
\begin{equation}
A = -1 + \frac{1}{2}(2+r^2)^2 = 1+2r^2 + \frac{1}{2}r^4,
\label{eq:app:A}
\end{equation}
and
\begin{equation}
u=x^4,
\end{equation}
we can write the quadrupled equation (\ref{eq:app:quad}) as
\begin{equation}
2A = u + \frac{1}{u}.
\label{eq:app:2A}
\end{equation}
Multiplication by $u$ and adding $A^2$ on each side gives
\begin{equation}
u^2 - 2Au + A^2 = -1 + A^2,
\end{equation}
from  which one can immediately read off the two solutions of eq.~(\ref{eq:app:2A}) as
\begin{equation}
u =  -A \pm \sqrt{A^2-1}.
\end{equation}
Given (\ref{eq:app:Hr}), the difference between the two roots,
\begin{equation}
u_1 - u_2 = 2\sqrt{A^2-1},
\end{equation}
provides an expression for the nominator of $\mathcal{H}_r$ in (\ref{eq:app:Hr}).
With the two roots, we can also rewrite the quadratic equation in the
form
\begin{equation}
u^2-2Au + 1 = (u-u_1)(u-u_2),
\end{equation}
and if we now set $u=1$, we obtain
\begin{equation}
2(1-A) = (1-u_1)(1-u_2),
\end{equation}
which gives us an expression for the denominator of $\mathcal{H}_r$ in (\ref{eq:app:Hr}).
Combining the two expressions, as Einstein did on p.~[48], we obtain
\begin{eqnarray}
\mathcal{H}_r &=& \sqrt{\frac{A+1}{A-1}}\\
              &=& \sqrt{1 + \frac{1}{r^2\left(1+\frac{1}{4}r^2\right)}},
	      \label{eq:app:Hrfinal}
\end{eqnarray}
where we have inserted (\ref{eq:app:A}) to obtain the second line.

\begin{figure}
\hskip -1.0cm
\includegraphics[scale=0.6]{fig5a}
\vskip -9.0cm
\hskip 4cm
\includegraphics[scale=0.6]{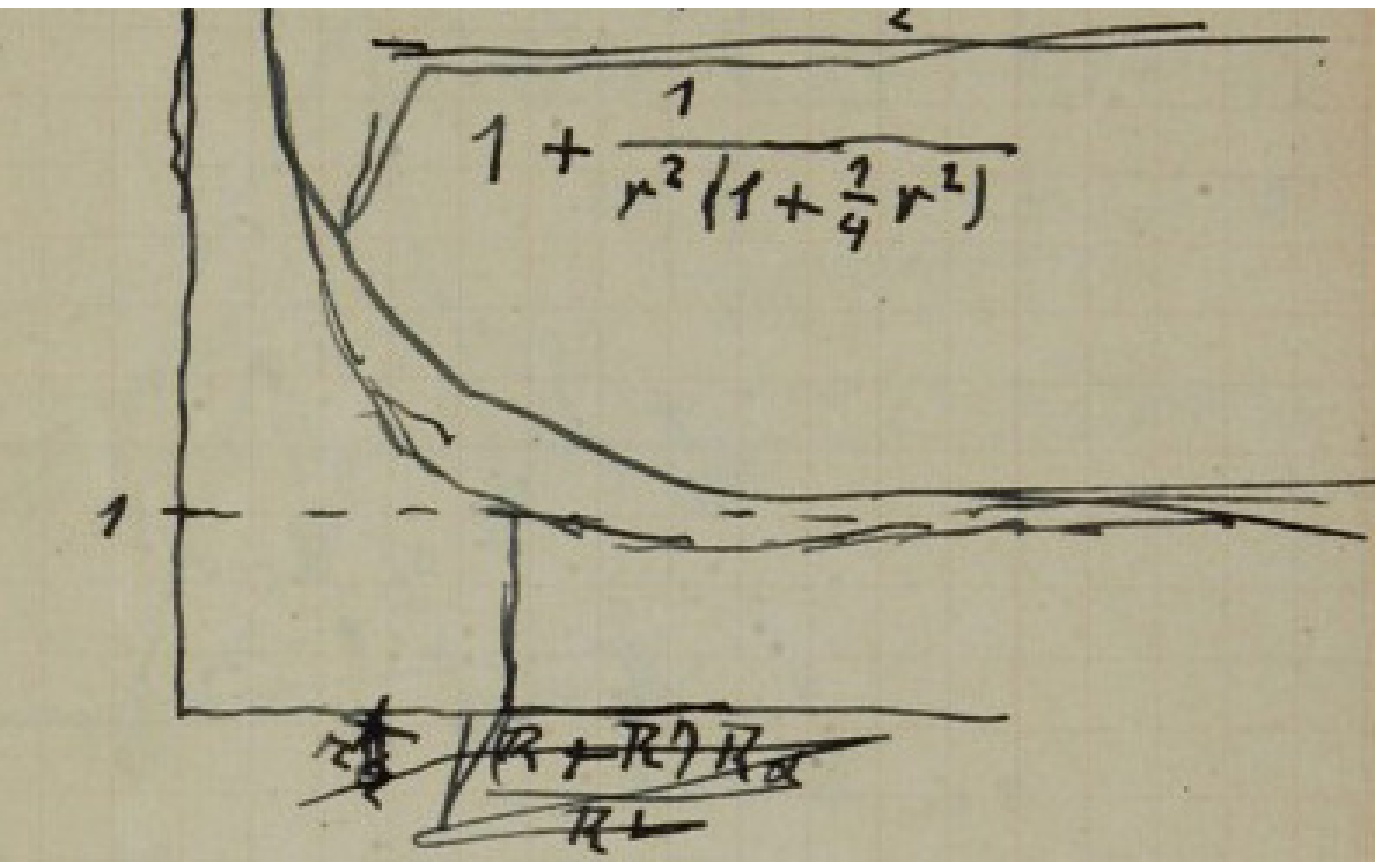}
\vskip 5cm
\caption{A plot of the expression (\ref{eq:app:Hrfinal}) for the relative brightness
	$\mathcal{H}_{\rm r}$ as a function of $r$. The inset is from \cite[p.~587]{CPAE3}.}
\label{fig:Hr}
\end{figure}

We now have an explicit expression for the relative brightness as
a function of the dimensionless variable $r$. We now evidently see that
$\mathcal{H}_r\rightarrow 1/r$ for $r\rightarrow 0$, and that $\mathcal{H}_r$
approaches $1$ asymptotically from above for large $r$, see Fig.~(\ref{fig:Hr}). 

Let us now reconstruct Einstein's order-of-magnitude estimate for the expected frequency
of the phenomenon on p.~[45]. The explicit expression for the relative brightness gives us
a measure of the maximal distance $r$ for which significant magnification is
obtained. 
We can look at specific values of $\mathcal{H}_{\rm r}(r)$. For instance, for $r_0 = \frac{1}{2}$
we find
\begin{equation}
\mathcal{H}_r(\frac{1}{2}) = \sqrt{1+\frac{1}{\frac{1}{4}\left(1+\frac{1}{16}\right)}}
\approx \sqrt{5} \approx 2.
\end{equation}
Hence, Einstein concluded that up to a distance of $r_0=\frac{1}{2}$ one would obtain
an increase of the intensity by a factor of $2$.
In other words, if we write the intensity $\mathcal{H}_{\rm r_0}$ asymptotically for
small $r_0$ and $R'\gg R$ as
\begin{eqnarray}
\mathcal{I}_0 \frac{1}{r_0} &=& \mathcal{I}_0 \frac{1}{r} \sqrt{\frac{R'(R+R')\alpha}{R}}
\label{eq:app:I0} \\
&\approx& \mathcal{I}_0 \frac{R'}{r}\sqrt{\frac{\alpha}{R}}\qquad,
\end{eqnarray}
we see that for a lensing star at a distance of $R$, the relative increase in intensity
is given by 
\begin{equation}
\frac{r}{R'} = \operatorname{tg}\bar{\alpha}.
\end{equation}
Here $\bar{\alpha}$ is the angle that determines how well the distant star has to be aligned
with the lensing star and the observer to produce appreciable magnification. In order to get an
order-of-magnitude estimate for this angle, one needs an order-of-magnitude estimate for 
$\sqrt{\frac{\alpha}{R}}$. In order to obtain such an estimate, Einstein notes that the ratio
of the solar Schwarschild radius $\alpha$ to the solar equatorial radius $R_s$ is 
given approximately by
\begin{equation}
\frac{\alpha}{R_s} = 3\cdot 10^{-6}.
\label{eq:num1}
\end{equation}
The radius of the sun is $2$ light seconds, and the distance of the nearest stars is of the
order of $10$ light years, or
\begin{equation}
10^5\cdot 365 \cdot 10 \approx 4\cdot 10^8 \quad {\rm light seconds}.
\label{eq:num2}
\end{equation}
It follows that $\frac{\alpha}{R}$ for a star of $1$ solar mass $10$ lightyears away
is
\begin{equation}
\frac{\alpha}{R} = \frac{\alpha}{R_s}\cdot\frac{R_s}{R} \approx 10^{-14} \quad
{\rm or} \quad \sqrt{\frac{\alpha}{R}} \approx 10^{-7}.
\end{equation}
To see the distant star with double intensity, we therefore have
\begin{equation}
2 = \frac{10^{-7}}{\operatorname{tg}\bar{\alpha}},
\end{equation}
so that the angle $\bar{\alpha}$ is of order $10^{-7}$.  A linear angle corresponds to
a solid angle roughly by taking its square. Thus, the angular size of the region where the
distant star needs to be found behind a massive star in order to be magnified in the lens
is of order $10^{-14}$.  In angular units, the total sky has an area of $4\pi\approx 10$,
so that the angular size of the region in question covers a fraction of $10^{-15}$ of the
total sky. This has to be contrasted with the average density of stellar population in the sky.
The {\em Bonner Durchmusterung} listed of the order of $3\cdot 10^{5}$ stars to ninth magnitude for 
the northern hemisphere, so a reasonable average density of the number of stars would be $1$ 
star per $10^{-5}$ of the sky.%
\footnote{On the relevant page under discussion here, we also find a little sketch by Einstein
	of a circle and the angle of its radius for a point some distance away. The precise
	meaning of this sketch is unclear but the numbers written next to it suggest
	that Einstein was considering the order of magnitude for the angular size of the moon. 
	The radius of the moon is seen under an angle of $15'$ from the earth, and the mean 
	distance between the earth and the moon in units of the lunar radius is about $200$,
	which translates to an angle of $50^{o}$.}

On the back of the loose sheet [p.~44] we find a few more calculations related to 
order-of-magnitude estimates that start from (\ref{eq:app:I0}). Einstein here again
goes back to the definition of $r_0$ and $\rho_0$ in terms of $R$, $R'$, and $\alpha$.%
\footnote{One can see here that Einstein corrected an error in his earlier calculations
	on [p.~43], where he had erroneously written the second term of the lensing 
	equation (\ref{eq:app:lensing})
	with $R$ instead of $R'$, which resulted in a confusion of the factors 
	of $R$ and $R'$ in expressions (\ref{eq:app:factor}) and (\ref{eq:app:r0}).}
Again, he observes that $r_0=\frac{1}{2}$ would give twice the usual intensity, and
rewrites (\ref{eq:app:r0}) for this case:
\begin{equation}
r = \frac{1}{2}\sqrt{\frac{R'(R+R')\alpha}{R}}.
\end{equation}
The latter equation for $R'\gg R$ turns into 
\begin{equation}
\frac{r}{R'} \approx \frac{1}{\sqrt{\alpha}{R}},
\end{equation}
and for $R\ll R'$ into
\begin{equation}
\frac{r}{R'} \approx \frac{1}{2}\sqrt{\alpha}{R'}.
\end{equation}
Einstein concluded that the smaller of the two distances $R$ and $R'$
determines the angle $\frac{r}{R'}$. In the top right corner of the page, Einstein
jotted down another order-of-magnitude calculation, which I do not fully understand. 
Apparently, he computed the distance of $100$ lightyears in terms of centimeters
\begin{equation}
3\cdot 10^{10}\cdot 3\cdot 10^{7}\cdot 10^{2}  \quad {\rm [cm]} \approx 10^{20} \quad {\rm cm}
\end{equation}

\begin{figure}
\centering
\includegraphics[scale=0.6]{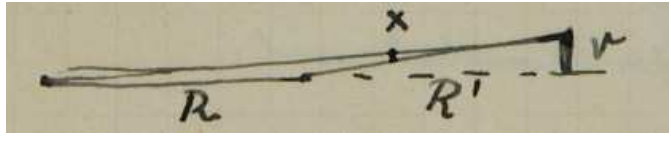}
\caption{A sketch in Einstein's scratch notebook to obtain eq.~(\ref{eq:x}). From \cite[p.585]{CPAE3}.}
\label{fig:sketch2}
\end{figure}

He also computed the angle $x$ under which the star at distance $R'$ and the star at distance
$R+R'$ would be seen by an observer at distance $r$ away from the connecting line between
the two stars if no lensing took place:
\begin{equation}
x = r\left\{\frac{1}{R'} - \frac{1}{R+R'}\right\} = r\frac{R}{R'(R+R')}.
\label{eq:x}
\end{equation}
The first equation can be read off from a little sketch of the geometry of light source,
lensing star, and observer, at the bottom of the page, see Fig.~(\ref{fig:sketch2}).

Let us finally comment on the calculations on pp.~[51] and [52]. As mentioned in the
main text of this article, Einstein here introduced a change of notation. On p.~[51],
he sketched again the geometry for stellar lensing. Here, the geometry has been turned by 90 degrees,
and the notation changed so that $R$ and $R'$ become $R_1$ and $R_2$, and $\rho$ and $r$
are interchanged to become $r$ and $\rho$, respectively. This change of notation is reflected in the
lensing equation, written down on p.~[52] as
\begin{equation}
\rho = r + R_1\left(w-\frac{\alpha}{r}\right) = 
\left(1+\frac{R_1}{R_2}\right)r - \frac{R_2\alpha}{r},
\label{eq:lensing2}
\end{equation}
where $\tan w = r/R_2$.
Einstein then immediately proceeded to compute
the magnification by taking the square of the lensing equation and then computing the 
derivative as
\begin{equation}
\frac{d(\rho^2)}{d(r^2)} = \left( 1+\frac{R_1}{R_2}\right)^2
-\frac{(R_1\alpha)^2}{r^4} = \mathcal{A}\cdot \frac{H}{H_0}.
\end{equation}
Instead of pursueing this calculation further, Einstein instead wrote
``apparent diameter of a Nova star,'' and  wrote down the solution of 
eq.~(\ref{eq:lensing2}) for $\rho=0$, as to obtain the diameter of
an Einstein ring:
\begin{equation}
r_0 = \sqrt{\frac{R_1R_2\alpha}{R_1+R_2}}.
\end{equation}
He computed the angle $w_0$ as
\begin{equation}
w_0 = \sqrt{\frac{R_1\alpha}{R_2(R_1+R_2)}}.
\end{equation}
The calculation ends with an attempt at a numerical order-of-magnitude estimation which seems
to proceed along the same lines as in eqs.~(\ref{eq:num1},\ref{eq:num2}). The calculation, 
however, was broken off, and the whole page was struck through.

\section*{Acknowledgments}

I wish to thank Diana Buchwald for a critical reading of an earlier version of this
paper, and Hilmar Duerbeck for some helpful comments.
Unpublished correspondence in the Albert Einstein Archives is quoted by kind permission.

\end{document}